\documentclass{ieeeaccess}
\usepackage{cite}
\usepackage{amsmath,amssymb,amsfonts}
\usepackage{graphicx}
\usepackage{textcomp}
\usepackage{array}
\usepackage{tabularx}
\usepackage{float}
\usepackage{multirow}
\usepackage{wrapfig}
\usepackage[version=4]{mhchem}
\usepackage{booktabs}
\usepackage{bm}
\usepackage{longtable}
\usepackage{lscape}
\usepackage{changepage}
\usepackage{algorithm}
\usepackage{algpseudocode}
\usepackage{enumitem}
\usepackage{enumerate}
\usepackage[table,xcdraw]{xcolor}  % Single load of xcolor
\usepackage{hyperref}
\usepackage{url}
\usepackage[utf8]{inputenc}
\usepackage{pifont} % For symbols
\usepackage{caption}

\hypersetup{
    colorlinks=true,
    citecolor={rgb:blue,1;red,0;green,0.5},  % Light blue color
    linkcolor=black,
    filecolor=magenta,
    urlcolor=black
}

\def\BibTeX{{\rm B\kern-.05em{\sc i\kern-.025em b}\kern-.08em
    T\kern-.1667em\lower.7ex\hbox{E}\kern-.125emX}}

\begin{document}
\history{\textbf{Received 25 July 2024; Accepted 30 September 2024; Date of publication TBD.}}
\doi{https://doi.org/10.48550/arXiv.2410.00012}

\title{Research on Enhancing C-V2X Communication via Danger-Aware Vehicular Networking}

\author{
    \uppercase{Lanre Sadeeq}\authorrefmark{1}\href{https://orcid.org/0009-0000-3075-8339}{\includegraphics[width=0.021\linewidth]{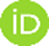}}, and
    \uppercase{Qasim Ajao}\authorrefmark{2}\href{https://orcid.org/0009-0007-7371-1742}{\includegraphics[width=0.021\linewidth]{fig1.png}} \IEEEmembership{Members, IEEE}
}
\address[1]{Department of Electrical and Computer Engineering, Georgia Southern University, Georgia, United States.}
\address[2]{Researcher at Department of Engineering, National Institute of Technology, West Africa Chapter, and Nigeria.}

\vspace{-15em}

\markboth
{Qasim \headeretal: Research on Enhancing C-V2X Communication}
{Qasim \headeretal: Research on Enhancing C-V2X Communication}

\corresp{\authorrefmark{1}Email: lanre07gbenga@ieee.org}

\begin{abstract}
This paper presents a protocol that optimizes message dissemination in C-V2X technology, which is crucial for advancing intelligent transportation systems (ITS) to enhance road safety. As vehicle density and velocity rise, the volume of data requiring communication significantly increases. By considering the danger and risk levels that vehicles encounter and using inter-vehicle proximity as a key indicator of potential hazards, the proposed protocol prioritizes communication, allowing vehicles facing higher risks to transmit their messages first. Our results show that this prioritization effectively reduces the number of concurrent transmissions, leading to improved performance metrics such as packet delivery ratio, throughput, latency, and lower probabilities of channel congestion and collision.
\end{abstract}


\titlepgskip=-33pt

\maketitle 

\section{INTRODUCTION}
The World Health Organization reports that car accidents cause over 1.2 million deaths and 50 million injuries each year \cite{b2}. Vehicle-to-Everything (V2X) communication can improve intelligent transportation systems (ITS), especially in making traffic safer \cite{b1}. V2X technology allows many autonomous vehicles to work together and perform tasks simultaneously, making it easier to merge, form platoons, and cross intersections \cite{b3}. Autonomous vehicles (AVs) and electric vehicles (EVs) are key to advancements in ITS, relying on electric propulsion, advanced batteries, sensors, processing, and AI technology. Efficient V2X communications within the ITS framework are crucial for AVs to improve traffic safety and control \cite{b4}. Protocols like Dedicated Short-Range Communications (DSRC) and Cellular Vehicle-to-Everything (C-V2X) are essential for effective communication between AVs, other vehicles, infrastructure, and pedestrians \cite{b5} \cite{b6}. This paper presents a strategy to resolve vehicle access conflicts by prioritizing messages based on the likelihood of collisions and filtering them to reduce network congestion. The adaptive V2X communications system dynamically responds to accidents by prioritizing signals according to collision risk. The success of this system depends on the widespread use of V2X-enabled AVs and the strength of the supporting communication infrastructure. 

\subsection{Related Research}
In ITS, radio communication is critical to traffic safety. DSRC and C-V2X are protocols used in autonomous cars for wireless communication \cite{b1} \cite{b6}. DSRC offers low-latency, dependable communication, essential for collision avoidance applications. However, by enabling direct communication between automobiles, infrastructure, and pedestrians, C-V2X leverages already-existing cellular networks to increase overall safety and efficiency \cite{b7}. Nevertheless, C-V2X struggles to achieve stringent latency requirements \cite{b8}. Future V2X obstacles cannot be resolved by DSRC or C-V2X alone, and combining the two might result in interference problems such as increased collisions from communication delays \cite{b6}. In vehicular communication, the size of the contention window (CW) significantly impacts network performance \cite{b9}. Although many approaches have been attempted to maximize CW size, performance is not always improved by making simple adjustments \cite{b10}. V2X networks need to function without the assistance of gateways. Reverse backoff mechanisms may enhance the chance of receipt by modifying CW values according to expiry; however, they must take into account the brief duration of cooperative awareness messages (CAMs), which need no backoff time to be sent. The behavior of DSRC networks is often examined using stochastic geometry \cite{b11}. The goal of this work is to enhance C-V2X communications for self-driving cars by maximizing back-off time allocation according to the "distance" factor.

\subsection{Impact Statement}
This paper presents a new strategy for vehicle networks to reduce the number of transmitters working at the same time. For effective V2X communication systems, the strategy considers factors like mobility, vehicle density, and competition for the spectrum. The goal is to improve real-world applications and reduce network congestion. The research has made significant progress in the following areas:
(1). Introducing resource allocation strategies based on danger assessment.
(2). Using the distance between vehicles' to determine the risk of collisions.

\section{MODEL CONFIGURATION}
This research aims to improve the efficiency of AV communication networks. It employs the IEEE 802.11 MAC strategy's Distributed Coordination Function (DCF) as its primary means of network access \cite{b12}. The study examines a given probability ($\tau$) for every station and computes the throughput (maximum amount of data transferred) and packet delivery rate (maximum number of successfully delivered packets; $PDR$) based on this probability using a two-dimensional Markov chain model. A two-dimensional stochastic process describes the behavior of every station. The countdown time until a station may attempt to communicate data again in this process is represented by $b(t)$, and the countdown stage at any given moment is represented by $s(t)$. 

Before transmitting data, every network node by the 802.11 DCF approach has to confirm that the channel is available. The node immediately begins transmitting data if the channel is available during the Distributed Interframe Space (DIFS). The initial node must wait a certain amount of time before attempting to use the channel again if it is currently being used by another node. The contention window (CW) size is the name given to this waiting period. Before it transmits data, the station selects a random number to count down from. This method adheres to a predetermined formula.

\begin{equation*}
\text{Back-off Counter} = \text{Random Value}() \times T_{\text{slot}} \tag{1}
\end{equation*}

where Random is an integer within the range \([0, CW]\), with \( CW \) varying between \( W_{\min} \) and \( W_{\max} \). The slot duration is represented by \( T_{\text{slot}} \). The most recent standard update includes these minimum and maximum contention window restrictions, especially in the PHY specifications section. When the channel is idle, the backoff counter begins to decrease; when the channel is busy, it stops. The station starts transmitting when the backoff counter reaches zero. This research uses a stochastic two-dimensional Markov chain model to assess the performance of the IEEE 802.11p DCF technique. To evaluate vehicle-to-vehicle communications, our study is predicated on certain assumptions. The model determines the transmission probability (\(\tau\)) by using the collision probability (\(P_{c}\)), as expounded in \cite{b6}. 

As an alternative, \cite{b13} offers the busy probability (\(P_{b}\)), which is ascertained using the stationary distribution \( b_{i, k} \), in which \( i \) denotes the backoff timing and runs from 0 to \( m \), the maximum backoff stage. When \( k \) approaches 0, a transmission event takes place. The possibilities of MAC transmission, busy, and collision are handled independently. When many cars (or stations) try to use the communication channel at the same time, a collision happens, and the busy probability (\(P_{b}\)) shows how likely it is that the channel will be filled.

The IEEE 802.11 DCF transmission method may be interpreted using a Markov chain to identify five different states, each of which represents a particular operational situation. When a channel is open for transmission and no other stations are using it, it is said to be in idle condition. A station may deliver its packet via the Successful transmission stage without running across any collisions \cite{b1} \cite{b6}. The Busy situation means that the channel is not accessible for new transmissions since another station is using it to broadcast. Collisions are identified at the first contention phase of transmission attempts, during the Collision state in the first stage (i) \cite{b13}. Collisions continue beyond the maximum number of re-transmission attempts (m) in the collision state at the maximum stage (m).

\begin{gather*}
P(i, k \mid i, k+1) = \frac{1 - P_b}{W_i},\quad i\in (0, m),\quad k\in(0, W_{i-2})\tag{2}\\
P(0, k \mid i, 0) = \frac{1 - P_c}{W_0}, \quad i \in (1, m), \quad k \in (0, W_{0-1})  \tag{3}\\
P(i, k \mid i, k) = \frac{P_b}{W_i}, \quad i \in (0, m), \quad k \in (0, W_{i-1})  \tag{4}\\
P(i, k \mid i-1, 0) = \frac{P_c}{W_i}, \quad i \in (1, m), \quad k \in (0, W_{i-1})  \tag{5}\\
P(m, k \mid m, 0) = \frac{P_c}{W_m}, \quad i = m, \quad k \in (0, W_{m-1}) \tag{6}
\end{gather*}

The back-off time counter's decline is seen in equation (1). Equation (2) illustrates how the counter for backoff time always resets to the '0' stage of backoff after a successful transmission. A random selection of the backoff time is made from the interval $\left(0, C W_{\min }\right)$ \cite{b6} if a transmission fails during the backoff stage $i$. 

The two-dimensional stochastic process $s(t), b(t)$ will be analyzed using the discrete-time Markov chain model. The limit of the probability that \( s(t) \) equals \( i \) and \( b(t) \) equals \( k \) as \( t \) approaches infinity is represented by \( b_{i, k} \), where \( i \) ranges from \( 0 \) to \( m \) and \( k \) ranges from \( 0 \) to \( W-1 \). The stationary distribution is shown by this limit. This stable distribution yields \( P_b \), which is the busy probability. 

We estimate the fundamental parameter \( \tau \) using the discrete-time Markov chain technique. Determining the throughput (\( S \)) and \( PDR \) requires this. For a single station, different states are indicated by the notation (\( i, k \)). To compute \( \tau \) \cite{b6}, several states, denoted by \( b_{i, k}, b_{0, 0}, b_{0, k}, b_{i, 0}, b_{m, 0}, \) and \( b_{m, k} \), are important for describing a single station. For the remaining states, the stationary probability \( b_{i, k} \) is specially employed.

\begin{gather*}
b_{0, k}=b_{0,0} \frac{1}{P_{b}-W_{0}}, \forall k \in\left(1, W_{0}-1\right)  \tag{7}\\
b_{i, 0}=P_{c}^{i} b_{0,0}, \forall i \in(1, m-1)  \tag{8}\\
b_{i, k}=b_{0,0} \frac{P_{c}^{i}}{1-\frac{P_{b}}{W_{0}}}\left(1-\frac{k}{W_{i}}\right), \forall i \in(1, m-1)  \tag{9}\\
b_{m, 0}=\frac{P_{m}}{1-P c^{i}} b_{0,0}  \tag{10}\\
b_{m, k}=b_{0,0}\left(1-\frac{k}{W_{m}}\right) \frac{1}{1-\frac{P_{b}}{W_{m}}} \frac{P_{c}^{m}}{1-P c}, \forall k \in\left(1, W_{m}-1\right) \tag{11}
\end{gather*}

As a result, the overall probability mass of these different probability states should equal 1. To get the remaining probability state $b_{0,0}$, also known as the likelihood state, we use the normalizing condition. The fact that $b_{0,0}$ is the unknown variable should be noted. So, in order to solve for $b_{0,0}$, we use the following normalizing condition:

\begin{equation*}
1=\sum_{i=0}^{m} \sum_{k=0}^{w_{i}-1} b_{i, k} \tag{12}
\end{equation*}

Once we determine $b_{0,0}$, it can be substituted into Equations (7) through (11) to ascertain the other probability states. As previously discussed, a transmission takes place when $k=0$. Hence, the likelihood of a station broadcasting may be expressed as follows:

\begin{equation*}
\tau=\sum_{i=0}^{m} b_{0,0} \tag{13}
\end{equation*}

The likelihood of at least one station broadcasting in a slot may be determined using the following formula:

\begin{equation*}
P_{t r}=1-(1-\tau)^{n} \tag{14}
\end{equation*}

Likewise, the likelihood of a packet communication being successful may be represented as:

\begin{equation*}
P_{s u}=\frac{n \times \tau \times(1-\tau)^{n-1}}{1-(1-\tau)^{n}} \tag{15}
\end{equation*}

Where $n$ represents the total number of competing stations simultaneously.

\section{PROPOSED STRATEGY}
In this paper, we consider a simple case with two lanes and two autos traveling in different directions. Fifty cars are arranged haphazardly along the two-lane road in this layout. Every car is fully aware of the route, the configuration of the roads, and other automobile features. The transmission range, without taking into account hidden nodes, is $1,000 \mathrm{~m}$. The packet load size of 1,100 bytes is the same for all vehicles. The distance data is used to calculate the number of transmitters at any given moment. 

The standard distance formula may be used to calculate the distance between two vehicles at positions $\left(x_{1}, y_{1}\right)$ and $\left(x_{2}, y_{2}\right)$:

\begin{equation*}
D=\sqrt{\left(x_{2}-x_{1}\right)^{2}+\left(y_{2}-y_{1}\right)^{2}} \tag{16}
\end{equation*}

Vehicles that fall below a certain threshold value are given precedence based on the available distance information between each automobile. The acceptable range for vehicle transmission is specified by the threshold distance. $200 \mathrm{~m}$, $400 \mathrm{~m}$, and $600 \mathrm{~m}$ are the threshold values used in the simulation. The suggested approach assigns a duty to each transmitter according to their distance information, hence reducing the total number of transmitters in a scenario with 50 cars. The approach for the simulation results is presented using the MATLAB program. Introduced is the notion of CAM messages, which enable the vehicle transmitting to forward or follow the message to the next vehicle if it is within the designated distance limit. The decision to transmit data is contingent upon the vehicle's closeness to the intended recipient, and the distance between vehicles also influences the value of $\mathrm{CW}$. The following procedures were used to create the approach, as shown in the flowchart in Figure 1:

\begin{figure}[ht]
\centerline{\includegraphics[scale=0.26]{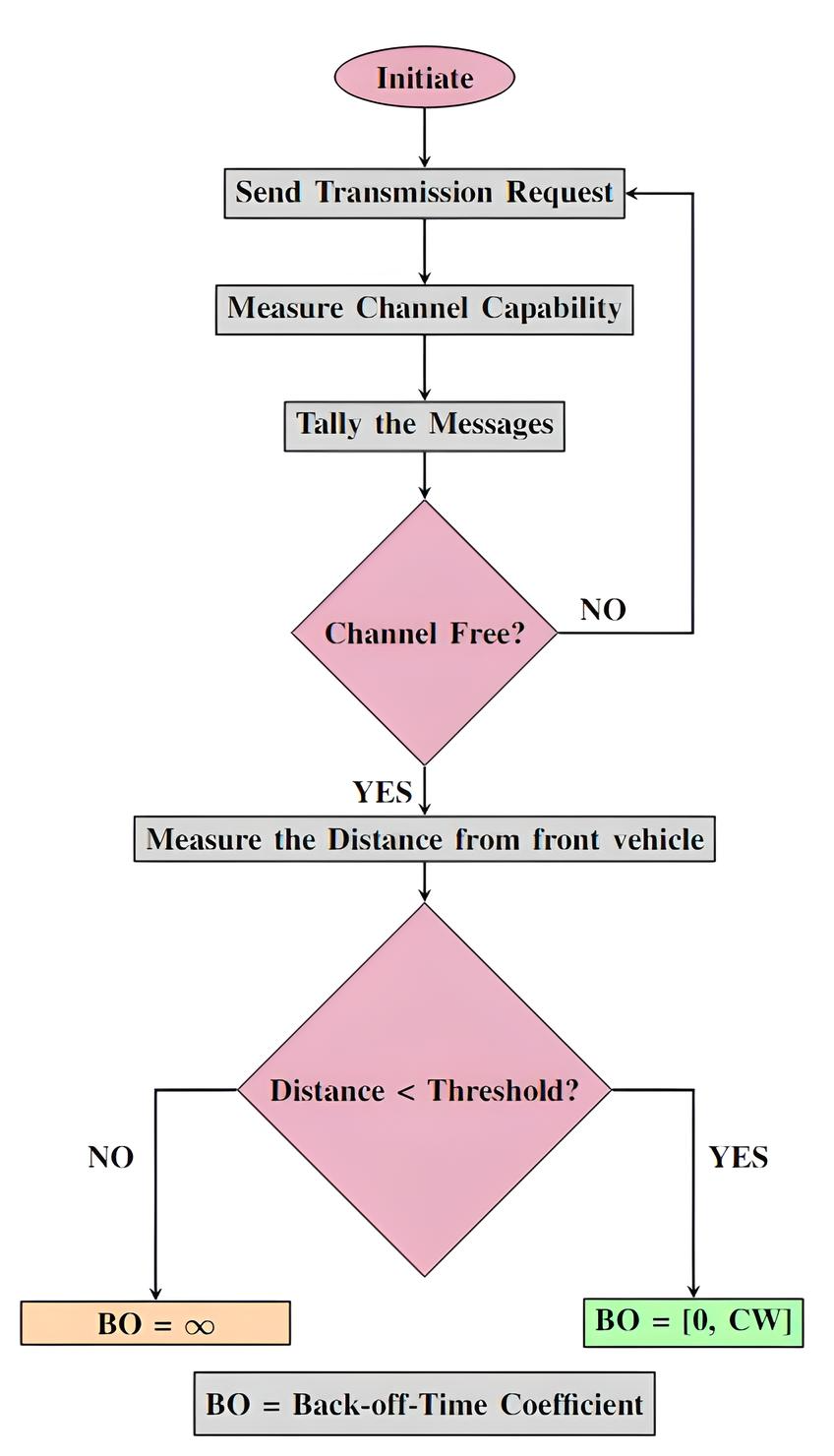}}
\caption{Rate for Packet Delivery (PDR)}
\end{figure}

\begin{itemize}
  \item Step 1: A linear section spanning a distance of 1,000 meters is lined with 50 automobiles, distributed equally and at random.
  \item Step 2: Calculate the separation between each pair of cars.
  \item Step 3: For simulations, a threshold value of 200, 400, and 600 meters is used.
  \item Step 4: We calculate a vehicle's risk by assessing how well it can keep a safe distance from other cars in relation to a predetermined threshold.
  \item Step 5: The car gets transmission authorization when its distance drops below a certain threshold.
\end{itemize}

\section{PERFORMANCE EVALUATION}
\subsection{PDR and Throughput}
A crucial metric that indicates the percentage of data successfully delivered relative to the overall amount of data transferred is the PDR. It is necessary for assessing system performance, together with throughput. In general, PDR and throughput both rise as transmissions fall. The backoff time drops to zero after the Distributed Inter-Frame Space (DIFS) phase, enabling packet transmission. The timer pauses until the channel is once again available if there is activity in the transmission medium. The term "busy likelihood" is often used to describe the duration of this delay. The probability of packet transmission ($\tau$) may be computed using the information on the likelihood of busyness and collisions \cite{b14}. A particular formula may be used to calculate throughput (S), a statistic that is often used to evaluate the performance of a system \cite{b6}.

\begin{equation*}
S=\frac{P_{s u} P_{t r} E[P]}{\left(1-P_{t r}\right)+P_{t r} P_{s u} T_{s}+P_{t r}(1-P s u) T_{c}} \tag{17}
\end{equation*}

Where \( P_{tr} \) transmission likelihood/probability, \( T_{s} \) successful transmission time, \( E[P] \) the average packet length, \( T_{c} \) collision time, \( \delta \) propagation delay, and \( P_{su} \) successful packet transmission likelihood/probability, are represented respectively.

\begin{gather*}
T_{s}=H+E[P]+S I F S+\delta+A C K+D I F S+\delta  \tag{18}\\
T_{c}=H+E[P]+D I F S+\delta \tag{19}
\end{gather*}

\subsection{Packet Delay}
The time it takes for a packet to travel from its origin to its destination is known as the delay. This comprises many time intervals, including the following: \(D_{\text{idle}}\) for idle delay, \(D_{\text{success}}\) for successful transmission delay, \(D_{\text{busy}}\) for busy delay, and \(D_{\text{collision}}\) for collision delay. Moreover, two critical probabilities impact the model's effectiveness: the likelihood that a vehicle will initiate a transmission (\(\tau_{\text{tr}}\)) and the likelihood that transmissions will overlap (\(\tau_{\text{nb}}\)). To ascertain the packet delay distribution at the MAC layer, one must consider the probabilities \(P_{\text{idle}}\), \(P_{\text{busy}}\), \(P_{\text{success}}\), \(P_{\text{own\_tx}}\), and \(P_{\text{collision}}\). These probabilities represent the likelihood that a station will try to broadcast, that a transmission will be successful, that a channel will be congested, and that a transmission collision will occur. The \(\tau\) value is matched with the values for \(\tau_{\text{tr}}\) and \(\tau_{\text{nb}}\).

\begin{gather*}
P_{\text{success}} = (n - 1) \left(\tau_{nb}\right) \left(1 - \tau_{tx}\right) \left(1 - \tau_{nb}\right)^{n-2} \tag{20}\\
P_{\text{idle}} = \left(1 - \tau_{tx}\right) \left(1 - \tau_{nb}\right)^{n-1} \tag{21}\\
P_{\text{collision}} = \tau_{tx} (n - 1) \left(\tau_{nb}\right) \left(1 - \tau_{nb}\right)^{n-2} \tag{22}\\
P_{\text{busy}} = 1 - P_{\text{idle}} - P_{\text{own\_tx}} - P_{\text{success}} - P_{\text{collision}} \tag{23}\\
P_{\text{own\_tx}} = \tau_{tx} \left(1 - \tau_{nb}\right)^{n-1} \tag{24}
\end{gather*}

The probability of each successful transmission may be used to portray the MAC layer delay as an ending renewal process. The sequence $S_{n}=T_{1}+T_{2}+T_{3}+\ldots \ldots... T_{n}+D_{s u c}$ may be used to represent this.

Throughput and PDR have increased as a consequence of the substantial transmission reduction achieved by our suggested approach. As seen in Figures 5 and 6, this tactic also reduces the likelihood of channel occupancy and accident rates. A reduction in the total time delay is facilitated by a drop in these variables as well as a shorter waiting period. Our results show that our strategy outperforms traditional baseline techniques. The sum of the transmission time (\(T_{tt}\)), total delay during collisions (\(T_{tc}\)), idle time (\(T_{emp}\)), and average backoff time (\(CW^*\) yields the total delay time (\(T_{td}\)).

\begin{gather*}
T_{\text{total delay}} = T_{\text{transmission time}} + T_{\text{collision time}} + CW^{} + T_{\text{empirical}} \tag{25}\\
T_{\text{transmission time}} = T_{\text{transmission slot}} \times N_{\text{transmissions}} \tag{26}\\
T_{\text{collision time}} = T_{\text{collision slot}} \times N_{\text{collisions}} \tag{27}\\
N_{\text{transmissions}} = P_{\text{transmission}} \times N_{\text{transmitters}} \tag{28}\\
N_{\text{collisions}} = P_{\text{collision}} \times N_{\text{transmitters}} \tag{29}
\end{gather*}

The variables $T_{t s p}$ and $T_{t s c}$ indicate how long it takes to send a single packet, how long a collision lasts overall, $T_{e m p}$ indicates when the network is idle, and $C W^{*}$ shows how long a device waits to send again after a collision.

\begin{equation*}
T_{t s p} = C T S + 3 S I F S + D I F S + \text{Data} + R T S + A C K \tag{30}
\end{equation*}

\begin{equation*}
T_{t s c}=R T S+D I F S \tag{31}
\end{equation*}

The typical delay period (also known as back-of-time) is represented by $C W^{*}$, as set out in the following equation:

\begin{equation*}
C W^{*}=\frac{C W_{\min } \times T_{\text {slot }}}{2} \tag{32}
\end{equation*}

The length of a slot time is represented by the variable $T_{\text {slot }}$, while the minimum size of the backoff window is indicated by the variable $C W_{\min }$ \cite{b6} \cite{b14}. As explained in Section IV, the equations for the probability of transmission, represented as \(P_{tr}\), and the likelihood of a collision, represented as \(P_{col}\), are given in Equations (14) and (23), respectively. The variable $P_{col}$ and the average number of transmitters may be used to compute the average number of collisions. Likewise, the same technique may be used to determine the average number of transmissions. Equations (25), (32), and (1) are used to calculate the total time delay. The shown graphic shows how the proposed method successfully reduces the total time delay with respect to the threshold distance.

\begin{figure}[ht]
\centerline{\includegraphics[scale=0.26]{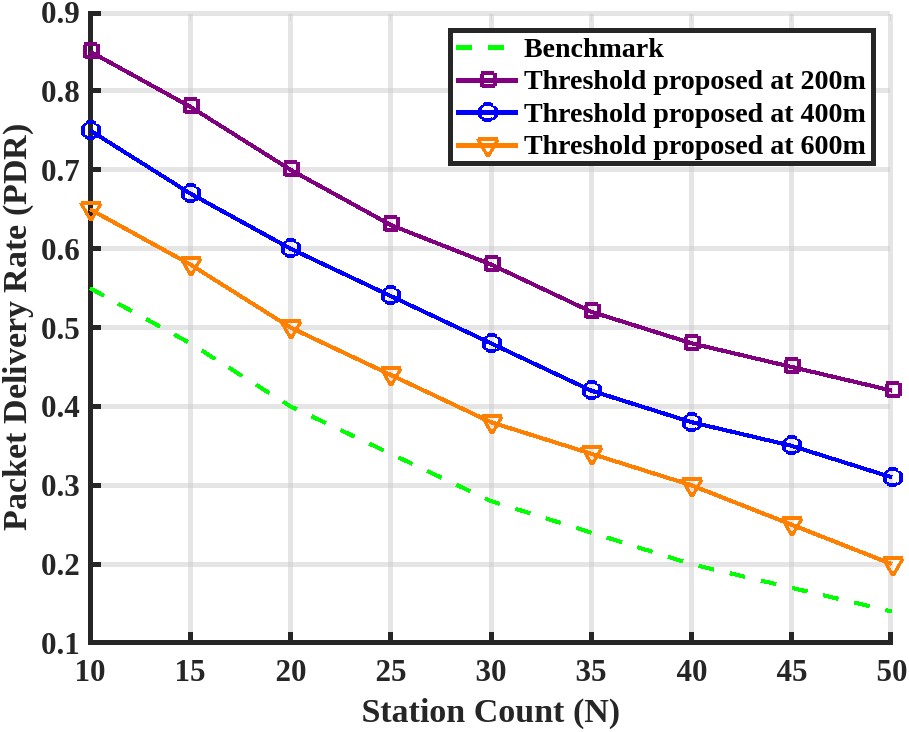}}
\caption{The Rate for Packet Delivery (PDR)}
\end{figure}

\begin{figure}[ht]
\centerline{\includegraphics[scale=0.26]{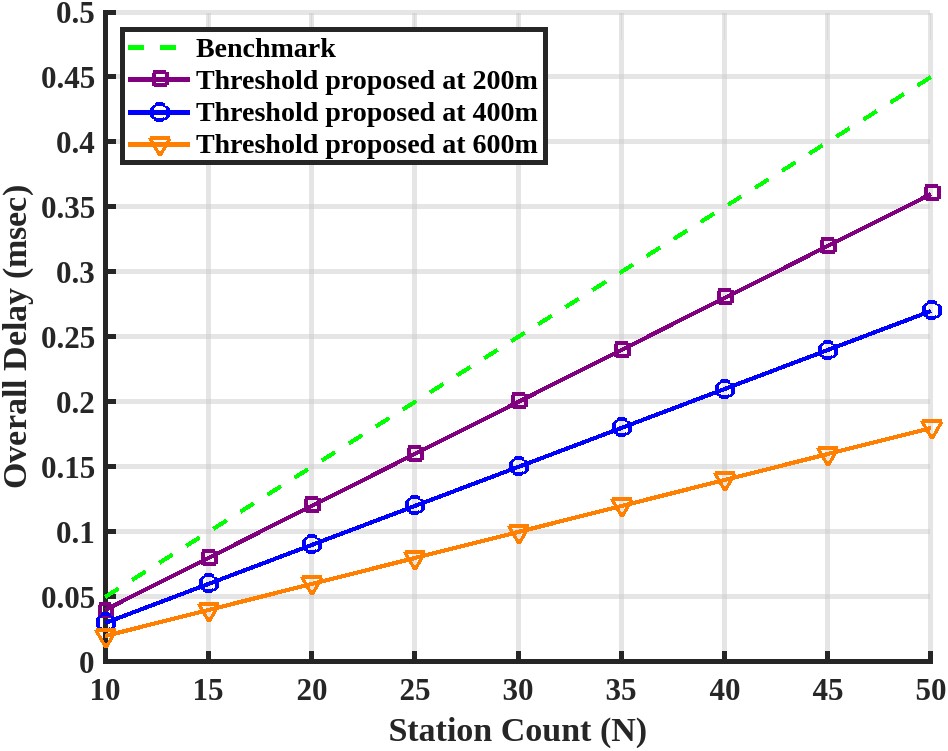}}
\caption{Overall/Total Delay}
\end{figure}

\begin{figure}[ht]
\centerline{\includegraphics[scale=0.26]{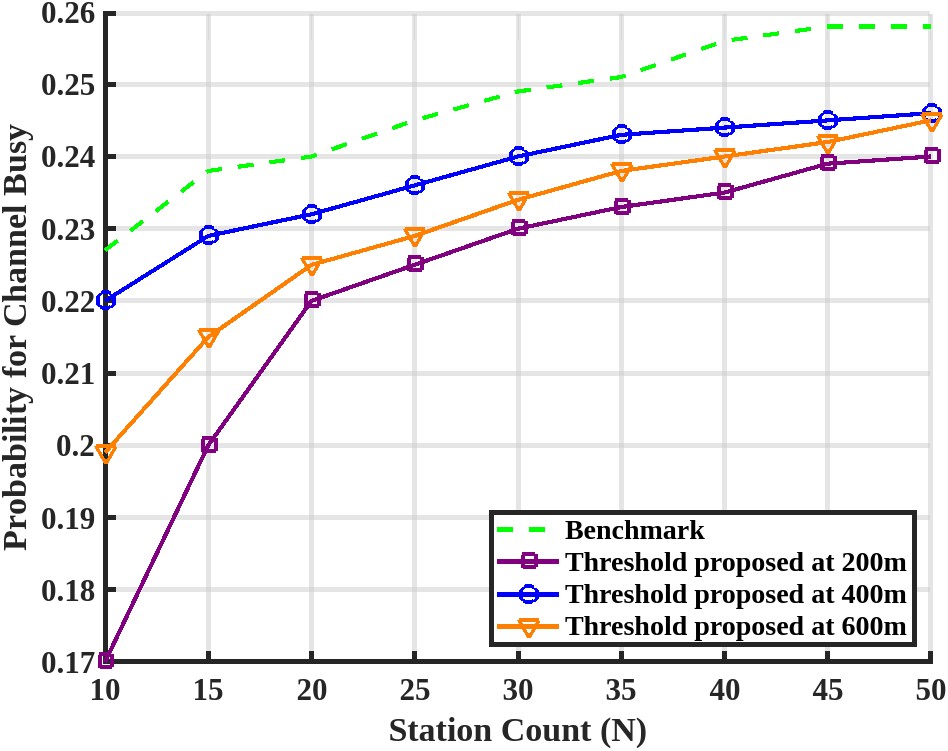}}
\caption{The Probability of Busy Channel}
\end{figure}

\begin{figure}[ht]
\centerline{\includegraphics[scale=0.26]{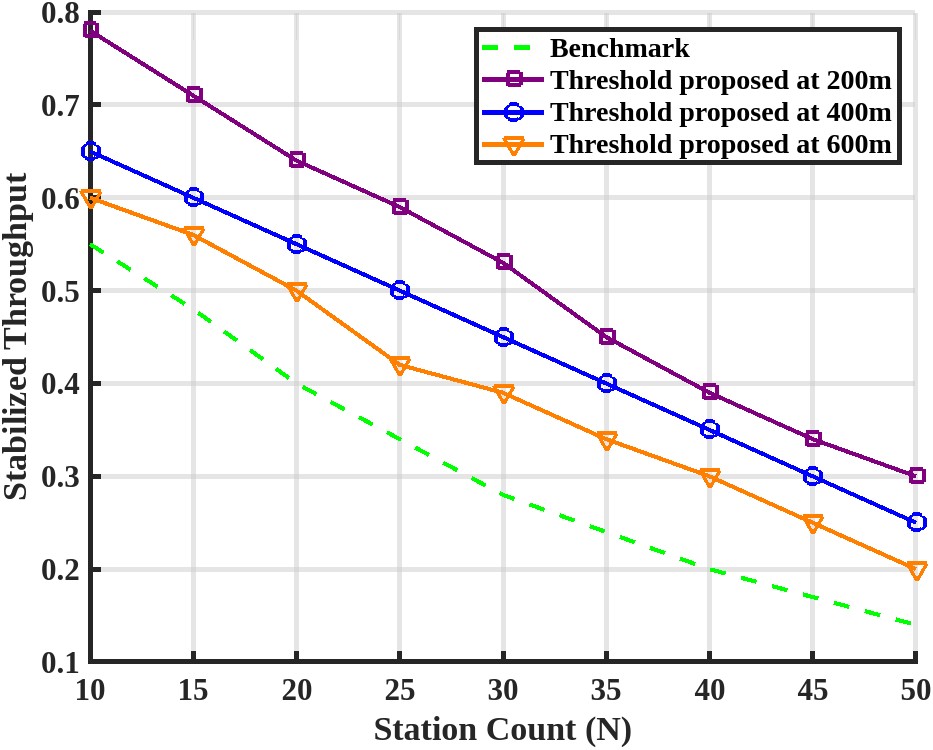}}
\caption{Stabilized/Normalized Throughput}
\end{figure}

\section{RESULTS}
The simulation's parameters are shown in Table 1. The findings demonstrate the effectiveness of our suggested strategy for managing car-to-car communication over a 1,000-meter route where individual vehicles are capable of sending messages. The PDR (packet delivery ratio) and throughput are two important performance indicators. These are shown by blue, red, and green dotted lines for distances of 200, 400, and 600 meters, respectively. The PDR and total delay are shown in Figures 2 and 3, which are crucial for assessing the efficacy of the technique. The results demonstrate that our approach outperforms the industry standard, particularly as the distance threshold drops. Throughput, or the amount of data transmitted successfully, is another important metric in addition to PDR. 

\begin{table}
 \caption{Simulation Parameters}
  \centering
  \begin{tabularx}{0.4\textwidth} { 
  | >{\raggedright\arraybackslash}X 
  | >{\centering\arraybackslash}X | }
%\begin{tabular}{|l|l|} 
\hline
Parameter & Value \\
\hline
DIFS & $65 \mu \mathrm{s}$ \\
SFIS & $35 \mu \mathrm{s}$ \\
Payload size & 1100 bytes \\
Propagation delay & $1 \mu \mathrm{s}$ \\
Slot duration & $15 \mu \mathrm{s}$ \\
$C W_{\min }$ & 7 \\
\hline
\end{tabularx}
\end{table}

Figure 5 illustrates how our approach outperforms the benchmark with a curve that resembles the PDR curve and is represented by blue, purple, and orange dots. Considerations such as transmission time, collision time, and time elapsed before attempting again are examined in the delay management study. Figure 4 illustrates how delays may be drastically reduced by lowering factors like the size of the data packets being transferred or the frequency of their transmission. Smaller data packets are less likely to collide and therefore simpler to transfer. Reducing the number of devices providing data helps minimize congestion, according to research on how full the network becomes at various packet reception levels and distances. Lowering the needs prior to initiating backups is also crucial for network optimization, avoiding channel overloads and needless channel use. By using these techniques, network traffic may be better managed and congestion issues can be resolved. Lowering the threshold also significantly lowers the likelihood of packet collisions, as seen in Figure 6. This validates the theory and demonstrates that future research should concentrate on enhancing these techniques and expanding their use.

\begin{figure}
\centerline{\includegraphics[scale=0.26]{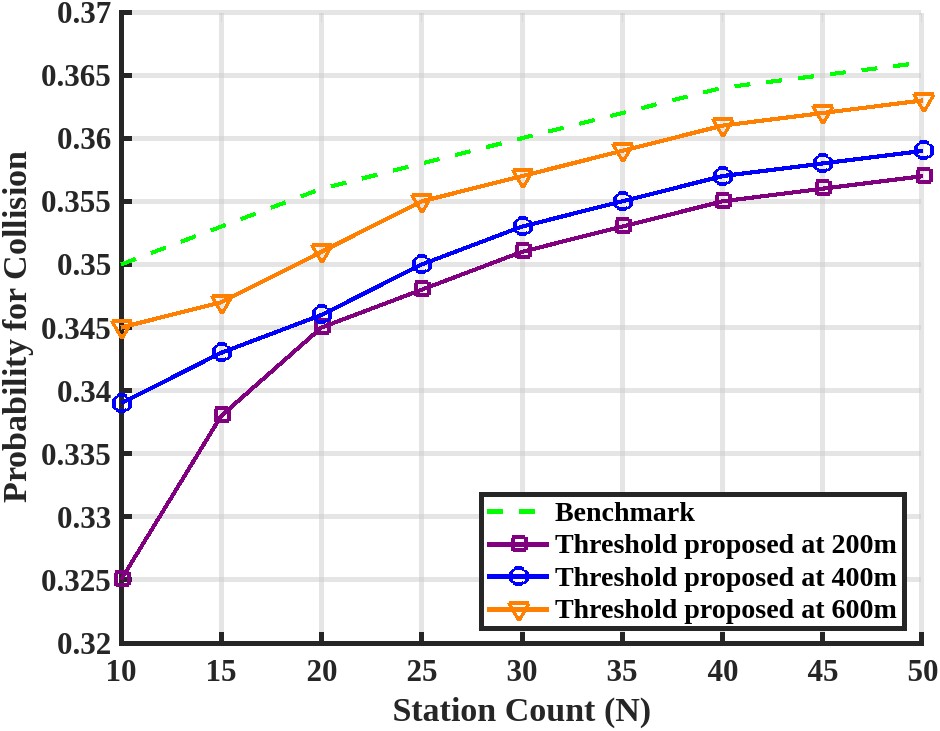}}
\caption{Probability for collision}
\end{figure}

\section*{CONCLUSION AND FUTURE WORK}
In this paper, we introduced an algorithm aimed at enhancing the performance of C-V2X communication networks through risk-level transmission allocation. The risk level was evaluated using inter-vehicle distance as a key metric. Our simulation results indicate that implementing this protocol can significantly boost the efficiency of C-V2X networks by prioritizing transmissions from vehicles at higher risk. Improvements were observed across various performance metrics, including packet delivery ratio, throughput, collision probability, channel occupancy, and overall latency. For future research, we have identified two main areas of focus. First, we plan to refine the proposed protocol to achieve a more nuanced resource allocation strategy. This will involve developing a more detailed classification of risk levels, allowing for finer adjustments in backoff times based on risk. Second, we aim to enhance the risk assessment by incorporating additional parameters beyond inter-vehicle distance. While this distance is a crucial indicator of potential hazards, other factors, such as driver behavior and environmental conditions, also contribute to crash risks. Our goal is to create a comprehensive risk assessment framework that integrates these diverse elements.

\section*{Declarations}
The authors have no competing interests to declare that are relevant to the content of this article.

\section*{Data Availability Statement}
All data analyzed during this study are included in the article.

\EOD

\end{document}